# Elucidating the cellular determinants of the end-systolic pressure-volume relationship of the heart via computational modelling

Francesco Regazzoni[1], Corrado Poggesi[2], Cecilia Ferrantini[2]

[1] MOX, Department of Mathematics, Politecnico di Milano, 20133 Milan, Italy
`francesco.regazzoni@polimi.it`
[2] Department of Experimental and Clinical Medicine, University of Florence, 50134 Florence, Italy

## ABSTRACT

The left ventricular end-systolic pressure-volume relationship (ESPVr) is a key indicator of cardiac contractility. Despite its established importance, several studies suggested that the mechanical mode of contraction, such as isovolumetric or ejecting contractions, may affect the ESPVr, challenging the traditional notion of a single, consistent relationship. Furthermore, it remains unclear whether the observed effects of ejection on force generation are inherent to the ventricular chamber itself or are a fundamental property of the myocardial tissue, with the underlying mechanisms remaining poorly understood. We investigated these aspects by using a multiscale in silico model that allowed us to elucidate the links between subcellular mechanisms and organ-level function. Simulations of ejecting and isovolumetric beats with different preload and afterload resistance were performed by modulating calcium and cross-bridge kinetics. The results suggest that the ESPVr is not a fixed curve but depends on the mechanical history of the contraction, with potentially both positive and negative effects of ejection. Isolated tissue simulations suggest that these phenomena are intrinsic to the myocardial tissue, rather than properties of the ventricular chamber. Our results suggest that the ESPVr results from the balance of positive and negative effects of ejection, respectively related to a memory effect of the increased apparent calcium sensitivity at high sarcomere length, and to the inverse relationship between force and velocity. Numerical simulations allowed us to reconcile conflicting results in the literature and suggest translational implications for clinical conditions such as hypertrophic cardiomyopathy, where altered calcium dynamics and cross-bridge kinetics may impact the ESPVr.

## INTRODUCTION

An essentially complete description of the mechanical dynamics of the heart is given by the changes in pressure and volume of the ventricles during each beat. The end-systolic pressure-volume (ESPV) relationship of the left ventricle (LV), obtained by connecting a series of pressure-volume points at end-systole in different cycles with different end-diastolic volumes and different afterload resistances, has been proposed as the best approach to defining cardiac contractility [1].

The first presentation of the ESPV relationships dates back to the end of the 19th century when Otto Frank reported a schematic diagram summarizing his experiments on the frog heart [2]. Frank studied different mechanical conditions of contraction (isovolumetric, isotonic with no afterload, and afterloaded beats) and adjusted the initial filling volume of the frog's ventricle to different levels. From Frank's schematic diagram (redrawn in Fig 3 by Sagawa [3], who published an English translation of Frank's original paper) it can be seen that the ESPV relationship of isovolumetric contractions is above that of isotonic contractions (with zero afterload) while that of afterloaded contractions is in between.

This means that the end-systolic pressure of the ejecting contractions is lower than that of the isovolumetric contractions initiated at the same end-systolic volume. Otto Frank thus showed that the relationship between end-systolic pressure and volume is strongly influenced by the mechanical mode of contraction.

Although Frank's results were later confirmed in various experimental models (for a review see [4]), many other studies, including those of Suga and Sagawa in the isolated dog heart [5], [6], concluded that the LV ESPV relationship was a single curve relatively insensitive to loading conditions. The slope of this single relationship and its position on the pressure-volume plane were modified only by changes in cardiac inotropism.

Nevertheless, it has long been known that shortening of striated muscle reduces force generation [7], [8] and, similarly, cardiac ejection has been shown to have a deactivating effect that reduces LV end-systolic pressure (e.g. [9], [10]). These observations and recent studies in isolated cardiac trabeculae that introduced the concept of an ESPV area rather than a single relationship [11] are consistent with Frank's original findings and support the idea that the ESPV relationship may depend on the mechanical mode of contraction.

To complicate these conflicting results, previous studies in isolated (e.g. [12]) and in situ hearts (e.g. [13]) and in isolated cardiac trabeculae (e.g. [14]) have shown that, under certain experimental conditions, ejection (and shortening) may have the opposite effect to that originally described by Frank. This means that the end-systolic pressure of afterloaded contractions may exceed that of isovolumetric contractions started at the same end-systolic volume. Both negative and positive effects of ejection appear to be basic properties of cardiac muscle and not due to properties of the ventricular chamber [14].

Mathematical models were used to investigate the mechanisms underlying the dependence of the cardiac ESPV (tension-length) relationship on the mechanical mode of contraction. These models have studied the influence of the sarcomere force-length dynamics, thin filament inactivation, $Ca^{2+}$-myofilaments interaction and cross-bridge (XB) cooperativity on the load dependence of the pressure-volume relations. These studies considered either isolated tissue conditions (e.g. [15], [16], [17], [18]), or simplified ventricular models, such as the spherical model (e.g. [14], [19]). However, the mechanisms that can modulate and even reverse the effects of ejection (shortening) on the end systolic pressure have not yet been identified.

The variations in sarcomere length (SL) between diastole and systole and how changes in SL correlate with the ESPV relationship are not easy to assess experimentally. Indeed, the working range of sarcomeres in vivo is still poorly known. Data obtained in Langendorff-perfused hearts [20] may not be able to replicate the in vivo conditions, with all the constrains of the organ in its pericardial cavity. Fukuda's lab [21] developed a high speed (100–frames per second), high resolution (20-nm) imaging system to map cardiac sarcomere dynamics in living mice. These authors found broad distributions for both diastolic SL (1.8-2.4) and systolic SL (1.3-2.1 µm).

In this work, we used a multiscale computational model of cardiac mechanics to investigate the factors influencing the ESPV relationship and to elucidate the links between subcellular mechanisms and organ-level function. To our knowledge, this is the first in silico investigation of the effects of shortening on the ESPV relationship using a geometrically detailed three-dimensional (3D) model of cardiac mechanics. In particular, 3D computational models typically neglect the force-velocity dependence due to numerical instabilities associated with this subcellular mechanism [22], [23], [24], [25], [26]. However, by using a recently proposed numerical stabilisation technique [27], we were able to highlight the role of the force-velocity in shaping the ESPV relationship. Thus, this work demonstrates for the first time – to our knowledge – that a single computational model can predict both positive and negative effects of ejection, depending on the parameters governing the kinetics of subcellular processes. Furthermore, we provide for the first time a mechanistic explanation for the observed phenomena based on the *in silico* investigation.

## METHODS
We considered an *in silico* model of human cardiac electromechanics, based on the iHEART Simulator [28], [29]. The model is inherently multi-scale and is based on a biophysically detailed description of both micro- and macro-scale dynamics and their coupling.

### Micro-scale electrophysiology model
To model cardiomyocyte electrophysiology, we used the ToR-ORd model [30] which was developed with the aim of providing a basis for accurate and reliable *in silico* models of cardiac electromechanics. The model has been extensively calibrated and validated by using human data [30]. In our simulations, to avoid artefacts due to the transient associated with the initial condition, we simulated 1000 beats to reach a limit cycle and then used the final state as the initial condition for the simulations.

### Micro-scale active force generation model
We considered the RDQ20-MF model [31], which describes the generation and regulation of active force in human cardiomyocytes. The model is based on an explicit description of the major regulatory and contractile proteins. The calcium-driven regulation of the thin filament takes into account the end-to-end interaction of tropomyosin units, thus reproducing the cooperative force-calcium curves. Although the explicit description of end-to-end interactions typically leads to prohibitive computational costs [32], thanks to the strategy proposed in [33], the model is formulated as a system of ordinary differential equations. This dramatically reduces the computational cost and allows the explicit representation of cooperative activation in multi-scale simulations. The model also accounts for length-dependent effects on the thin-filament activation, through an effective calcium-sensitivity of troponin that increases with the SL, and a maximum tension that depends on the filament overlap. The model has been calibrated using human data, and it is able to reproduce the steady-state force-calcium and force-length relationships, the length-dependent prolongation of twitches and increase in peak force, and the force-velocity relationship [31].

In this work, we investigated the effect of the XB kinetics, i.e. the ability of XBs to cycle rapidly, on the ESPV relationship. To do this, we simultaneously rescaled the XB attachment and detachment rates in the RDQ20-MF model by an appropriate constant, so that the duty-ratio, and hence the maximal force, was unaffected. When the XB rates were rescaled by a constant greater than 1, the kinetics were accelerated, whereas when the rescaling constant was less than 1, the kinetics were slowed.

### Multi-scale electromechanical model
For the simulation of the multi-scale electromechanics, we relied on the eikonal-reaction-mechanics model [34]. Specifically, the propagation of the action-potential in the cardiac tissue was modelled by the eikonal-diffusion model [35], [36], suitable for sinus rhythm conditions. This model was coupled to the ToR-ORd model, by applying an electric current, representing the diffusion current, to each point of the domain, starting from the activation time predicted by the eikonal-diffusion model. The calcium concentration obtained from the ToR-ORd was in turn fed to the RDQ20-MF model to calculate the active tension generated. The mechanical behaviour of the myocardium was modelled by the elastodynamic equations [37], within a hyperelastic framework using the nearly incompressible exponential constitutive law of [38]. The hyperelastic material law was supplemented with an active stress tensor incorporating the active tension calculated by the RDQ20-MF model. The local SL and SL shortening/lengthening rates were derived from the tissue deformation and fed back into the force generation model, thus incorporating the effect of tissue strain and strain rate on active force generation.

## Left ventricle simulations

We considered an LV geometry derived from the Zygote 3D heart model [39], representing an average healthy human heart reconstructed from a high-resolution computer tomography scan, from which we recovered the unstressed reference configuration using the inverse displacement algorithm [40]. The fibres distribution was generated using the Bayer-Blake-Plank-Trayanova algorithm [41], [42]. Electrical activation was induced by localised stimuli at three points of the endocardium which replaced the electric conduction system. On the epicardial surface, we considered spring-damper boundary conditions to account for the effect of the pericardium and surrounding organs, while at the ventricular base we considered energy-consistent boundary conditions [43]. We applied pressure boundary conditions to the endocardium, with the pressure determined in different ways depending on the simulation protocol considered.

We considered two simulation protocols, namely isovolumetric beats and ejecting beats. In isovolumetric beats, to reproduce the experimental condition of clamped valves, we solved the elastodynamic equations under a constant volume constraint. In this case, the blood pressure was determined as the Lagrange multiplier that enforces the constraint. Instead, during ejecting beats the valves were kept closed until a pressure of 80 mmHg was reached, and then the LV model was coupled with a two-element Windkessel afterload model as described in [16] (see Figure 1a). In both cases, a limit cycle was reached to avoid artefacts due to initialization. Both protocols were run for different end-diastolic volumes (EDV), different afterload resistances ($R_{ar}$), different XB kinetics, and different calcium transients (CaT). Isovolumetric ESPV curves were obtained by connecting the peak pressure points obtained with isovolumetric beats for different EDVs (see Figure 1d).

## Isolated cardiac muscle simulations

To complement the simulations in the 3D LV geometry, we also performed simulations of isolated tissue. The aim was to loosely mimic the behaviour of individual cells during a heartbeat, but without the 3D chamber configuration, in order to investigate to what extent the results are tissue specific or organ-level phenomena, and to allow a more detailed investigation of the underlying mechanisms. Therefore, we considered a cardiomyocyte model described by the ToR-ORd model coupled with the RDQ20-MF model, but without spatial detail.

Again, two simulation protocols were considered, namely isometric twitches and shortening twitches (corresponding to the two protocols considered for the LV). In isometric twitches, the SL was kept at a constant a priori value throughout the simulation. Shortening twitches, on the other hand, were simulated with the setup shown in Figure 1o: the cardiac tissue was modelled by the parallel between a contractile element and a passive elastic element, connected in series with a viscous-elastic element, to mimic the presence of a load against which the muscle contracts. The length of the contractile element was kept constant until the sum of the active ($T_a$) and passive ($T_p$) tension associated with the cardiac tissue, denoted by $T$, reached a prescribed threshold $T_{th}$. From that moment on, the tissue was free to contract and then relax against the viscous-elastic series element. As we were not interested in the filling phase in this investigation, we did not include a restretch phase in the simulation.

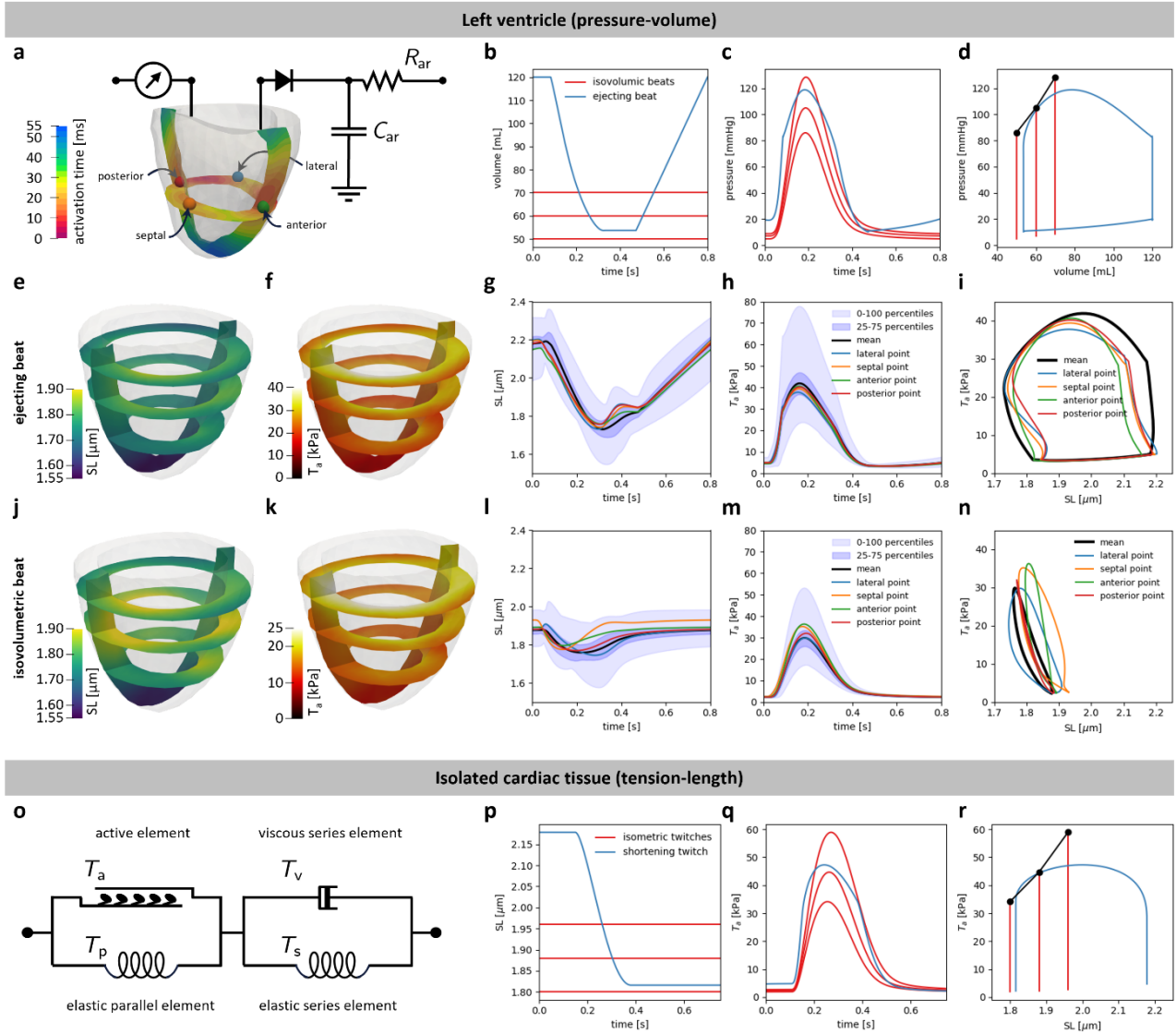

**Figure 1: Left ventricle model and Isolated cardiac tissue model.**

**a**: Illustration of the 3D left ventricle model, coupled to a Windkessel afterload model, characterised by arterial resistance $R_{ar}$ and arterial compliance $C_{ar}$. The long-axis and short-axis views show the activation times obtained under baseline conditions. Four representative points (septal, lateral, anterior, posterior) are shown to serve as reference markers in the upcoming illustrations. **b-c-d**: Pressure and volume traces of one ejecting beat (starting with EDV = 120 mL) and of three isovolumetric beats (with V = 50, 60, 70 mL). In the pressure-volume plane (d), the isovolumetric peaks are connected, showcasing the procedure followed to construct the isovolumetric ESPVr. **e**: Sarcomere length distribution at time t = 0.3 s (corresponding to the end of systole) for the ejecting beat considered in b-c-d. One long-axis and three short-axis sections are displayed to highlight the transmural distribution. **f**: Active tension distribution at time t = 0.3 s for the ejecting beat considered in b-c-d. **g-h**: Time traces of sarcomere length and active tension for the ejecting beat considered in b-c-d. The light shaded area represents the range between the minimum and maximum values over space, while the dark shaded area shows the range between the first and third quartiles. Black lines represent the space averages, while coloured lines correspond to the four representative points displayed in a. **i**: Loops in the tension-length plane (ejecting beat) associated with average values over space (black line) and with the four representative points displayed in a. **j-k-l-m-n**: same as for e-f-g-h-I, but this time considering the isovolumetric beat with EDV = 60 mL represented in b-c-d. Note that, while in e and j the same colour bar is employed to represent the sarcomere length, in f and k a different range of active tension is considered, due to the highly different tension peaks reached. **o**: Representation of the single cell model, coupled with a parallel elastic element and with a viscoelastic series element. **p-q-r**: tension and length traces of a shortening twitch (starting from SL = 2.18 μm) and three isometric twitches (with SL = 1.80, 1.88, 1.96 μm). In the tension-length plane (r), the isometric peaks are connected, showcasing the procedure followed to construct the isometric tension-length relationship.

The constitutive behaviour of the parallel passive element was modelled by the law derived from human tissue experiments in [44], i.e. $T_p(\lambda) = a(e^{b\lambda} - 1)$, where $\lambda$ is the tissue strain in the fibre direction. The tension generated by the viscous series component was modelled as $T_v = -\sigma \dot{\lambda}$, where $\sigma$ is the associated viscosity, while the elastic series component generated a tension equal to $T_s = k_s(\bar{\lambda} - \lambda)$, where $k_s$ is the series stiffness and $\bar{\lambda}$ is the strain corresponding to zero force. In summary, the model under consideration gives rise to the following differential equation:

$$\sigma \dot{\lambda} + k_s \lambda + T_p(\lambda) + T_a = \bar{T},$$

where $\bar{T}$ is a constant that depends on the above parameters. We calibrated the parameters to obtain tension-length transients similar to those obtained for single cardiomyocytes in LV simulations under baseline conditions (see Figure 1), thus obtaining $\sigma = 15 \text{ kPa} \cdot \text{s}$, $k_s = 25 \text{ kPa}$, $T_{th} = 30 \text{ kPa}$.

Similar to the LV simulations, both the isometric and shortening simulation protocols were run for different $SL_{ED}$ (end-diastolic SL), different viscosities $\sigma$, different XB kinetics, and different CaTs. Isometric end-systolic tension-length curves were obtained by connecting the peak tension points associated with isometric twitches for different $SL_{ED}$ (see Figure 1r).

## Quantification of shortening activation/deactivation

To quantify the contractile capacity and the effect of shortening on the force generated, we used different metrics, for both LV and for isolated muscle simulations (see Figure 2). For LV simulations, we define $V_{80}^{\text{ejec}}$ as the LV volume at which the pressure returns to the value of 80 mmHg after exceeding it at the beginning of systole in ejecting beats. Similarly, we define $V_{80}^{\text{iso}}$ as the LV volume at which the isovolumetric ESPV curve reaches a pressure of 80 mmHg. Obviously, the lower the $V_{80}^*$, the more effective the contraction. To quantify the effect of shortening on contractility, we define $\Delta V_{80} = V_{80}^{\text{iso}} - V_{80}^{\text{ejec}}$: a positive (respectively, negative) $\Delta V_{80}$ denotes the presence of shortening activation (respectively, deactivation).

We note that indices such as $V_{80}^{\text{ejec}}$ are often used in the clinical practice to quantify contractility [45]. However, from case to case, the largest volume shift between the isovolumetric and ejecting case may occur in correspondence of different pressure levels, and not necessarily at $p_{LV} = 80$ mmHg. Based on this observation, we define the following two additional metrics:

$$\Delta V_{\max} = \max_t \left( V^{\text{iso}}(p_{LV}(t)) - V_{LV}(t) \right),$$

$$\Delta p_{\max} = \max_t (p_{LV}(t) - p^{\text{iso}}(V_{LV}(t))),$$

where $V_{LV}(t)$ and $p_{LV}(t)$ are the LV volume and pressure transients obtained from an ejecting beat. $V^{\text{iso}}(p)$ denotes the LV volume for which the peak isovolumetric LV pressure is equal to $p$, while $p^{\text{iso}}(V)$ denotes the peak isovolumetric LV pressure obtained with an LV volume of $V$. In other words, $\Delta V_{\max}$ and $\Delta p_{\max}$ correspond to the maximum leftward and upward shifts, respectively, in the pressure-volume planes between the isovolumetric ESPV curve and the ejecting pressure-volume loop (see Figure 2). Similar to $\Delta V_{80}$, positive values of both $\Delta V_{\max}$ and $\Delta p_{\max}$ indicate an activating effect of shortening, while negative values indicate a deactivating effect. The more positive the value, the stronger the activation; the more negative the value, the stronger the deactivation.

For isolated muscle simulations, we define analogous metrics, by replacing volume by SL, and pressure by total tension $T = T_a + T_p$. Specifically, we define $SL_{40}^{\text{short}}$ and $SL_{40}^{\text{iso}}$ as the SL associated with $T = 40$ kPa in shortening and isometric twitches, respectively, and we define $\Delta SL_{40} = SL_{40}^{\text{iso}} - SL_{40}^{\text{short}}$ as a measure of shortening activation. Then, analogous to above, we define

$$\Delta SL_{\max} = \max_t \left( SL^{\text{iso}}(T(t)) - SL(t) \right),$$

$$\Delta T_{\max} = \max_t (T(t) - T^{\text{iso}}(SL(t))),$$

where $SL^{\text{iso}}(T)$ and $T^{\text{iso}}(SL)$ denote the isometric length-tension and tension-length curves.

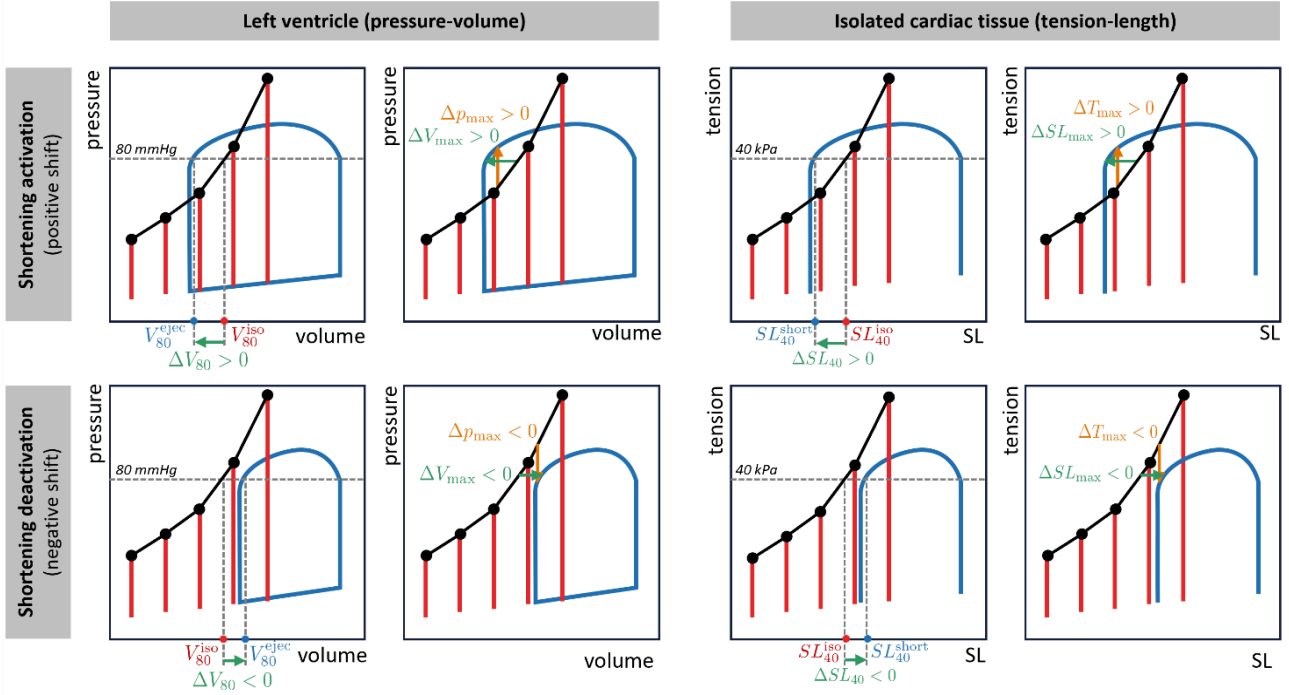

**Figure 2: Representation of the metrics used to quantify the impact of shortening upon contractility.**

Left: metrics used in LV simulations (pressure-volume plane); right: metrics used in isolated tissue simulations (tension-length plane). The first row represents the case of a positive effect of shortening on contractility (shortening activation), while the second row corresponds to the case of a negative effect (shortening deactivation). For the LV (or isolated tissue) case, consistent with Figure 1d and Figure 1r, the blue lines and labels correspond to ejecting beats (or shortening twitches), the red lines and labels correspond to isovolumetric beats (or isometric twitches), while the black lines represent the end-systolic pressure-volume (or tension-length) relationship. Green labels refer to metrics related to volume (or length), namely $\Delta V_{80}$, $\Delta V_{\max}$, $\Delta SL_{40}$, and $\Delta SL_{\max}$. Orange labels refer to metrics related to pressure (or tension), namely $\Delta p_{\max}$ and $\Delta T_{\max}$.

## Numerical approximation and simulations

To simulate the isolated muscle model, we used the staggered time stepping scheme described in [46], where the thin filament activation variables are updated by an explicit probability-preserving scheme, while the XB variables are updated by an implicit scheme with a coarser time step size. This scheme provides a very good balance between numerical accuracy and computational cost [46]. For shortening twitches, the parallel-series elements were coupled to the contractile element through an implicit scheme, relying on the Newton method to solve the associated nonlinear system of equations. Due to the lack of feedback from the contractile model to the ionic model, the solution of the ToR-ORd model was precomputed, using an adaptive Runge-Kutta method, and the obtained CaT was provided as input to the RDQ20-MF model.

For the LV model, we performed the spatial discretization using bilinear Finite Elements on hexahedra, and relied on the time-staggered scheme presented in [34] to couple the submodels. Specifically, the solution of the ToR-ORd model was precomputed as described above, and the limit cycle was used in the 3D simulation. The micro-scale activation model and the macro-scale mechanics model were coupled using the stabilised scheme recently proposed in [27]. Remarkably, this scheme allowed, for the first time, to remove the numerical oscillations that arise in staggered schemes when the force-velocity dependence is included in the model, thus allowing the study of the organ-level effects of this regulatory mechanism, which – as will be shown later – plays a key role in this work. The simulations were performed on the MOX (Dipartimento di Matematica, Politecnico di Milano) cluster computer, using life[x] [47], [48], a high-performance computing C++ library for finite element simulations of the cardiac function.

## RESULTS

### Baseline conditions

Figure 3 (middle column, labelled "XB kinetics x1") shows the results obtained with the LV model under baseline conditions. The ejecting beats were performed with three different values of EDV (100, 120, 140 mL) and three different values of arterial resistance ($R_{ar}$ = 36, 48, 60 MPa · s m$^{-3}$). As a matter of fact, the envelope of all the ejecting pressure-volume loops obtained is close to the isovolumetric ESPV curve. However, the two curves do not match perfectly: for large EDV, we observe a slight shortening activation effect, as the ejecting pressure-volume loops are slightly above the isovolumetric curve; for small EDV, on the other hand, they fall slightly below the isovolumetric curve, indicating shortening deactivation. Consistently, both $\Delta V_{\max}$ and $\Delta p_{\max}$ are slightly positive (respectively, negative) for large (respectively, small) EDV.

It is noteworthy that the ejecting pressure-volume loops are not tangent to a single curve, but rather define an area in the pressure-volume plane. This effect is most noticeable for volumes below 70 mL, where multiple curves reach different pressures depending on the preload.

### The impact of XB kinetics

We investigated the impact of modulating the XB kinetics on heart contractility. Figure 3 compares the results obtained under baseline conditions (XB kinetics x1), with the cases where XB rates were accelerated by a factor of 2 (XB kinetics x2) or slowed down by a factor of 2 (XB kinetics x0.5). The results show that an increased (or decreased) XB kinetics has a clear shortening activation (or deactivation) effect. The same conclusion is supported by the trend of $\Delta V_{80}$, $\Delta V_{\max}$ and $\Delta p_{\max}$. In addition, in the second row of Figure 3, we report the space-averaged active tension and the average SL post-processed from both ejecting and isovolumetric beat simulations. These results show that the average tension-length traces closely resemble the corresponding pressure-volume traces: the shortening activation/deactivation effect of XB kinetics is already apparent at the cellular level.

To characterise the impact of XB kinetics at the tissue level, we report in Figure 3d the normalised tension-length curves obtained on isolated tissue simulations when XB rates are rescaled by factors of 0.5, 1 and 2. The figure clearly shows that the faster the XB kinetics, the greater the ability of the tissue to sustain high shortening velocities.

In Figure 4, we report the results of numerical tests similar to those in Figure 3 but performed on isolated tissue. Specifically, we compared isometric end-systolic tension-length curves with those obtained from shortening twitches. Shortening twitches were performed by considering three values of $SL_{ED}$ (2.0779, 2.1783, 2.2765 μM, chosen by taking the mean end-diastolic SL in the LV tests reported in Figure 3), and three different viscous damping coefficients, obtained by starting from the baseline value $\sigma = 15$ kPa · s with steps of 3,75 kPa · s. The results are in good agreement with those obtained at the chamber level. Specifically, an increase in diastolic SL induced shortening activation (slightly more pronounced than in LV tests), and an increase in XB kinetics had a marked shortening activation effect.

Remarkably, in the case of fast XB kinetics, it is even more apparent that connecting the end-systolic points obtained by varying the preload and afterload resistances does not give a unique relationship but rather an area. Similarly, in isolated tissue simulations, the end-systolic tension-length relationship defines a region in the tension-length plane, with a larger area when XBs are accelerated.

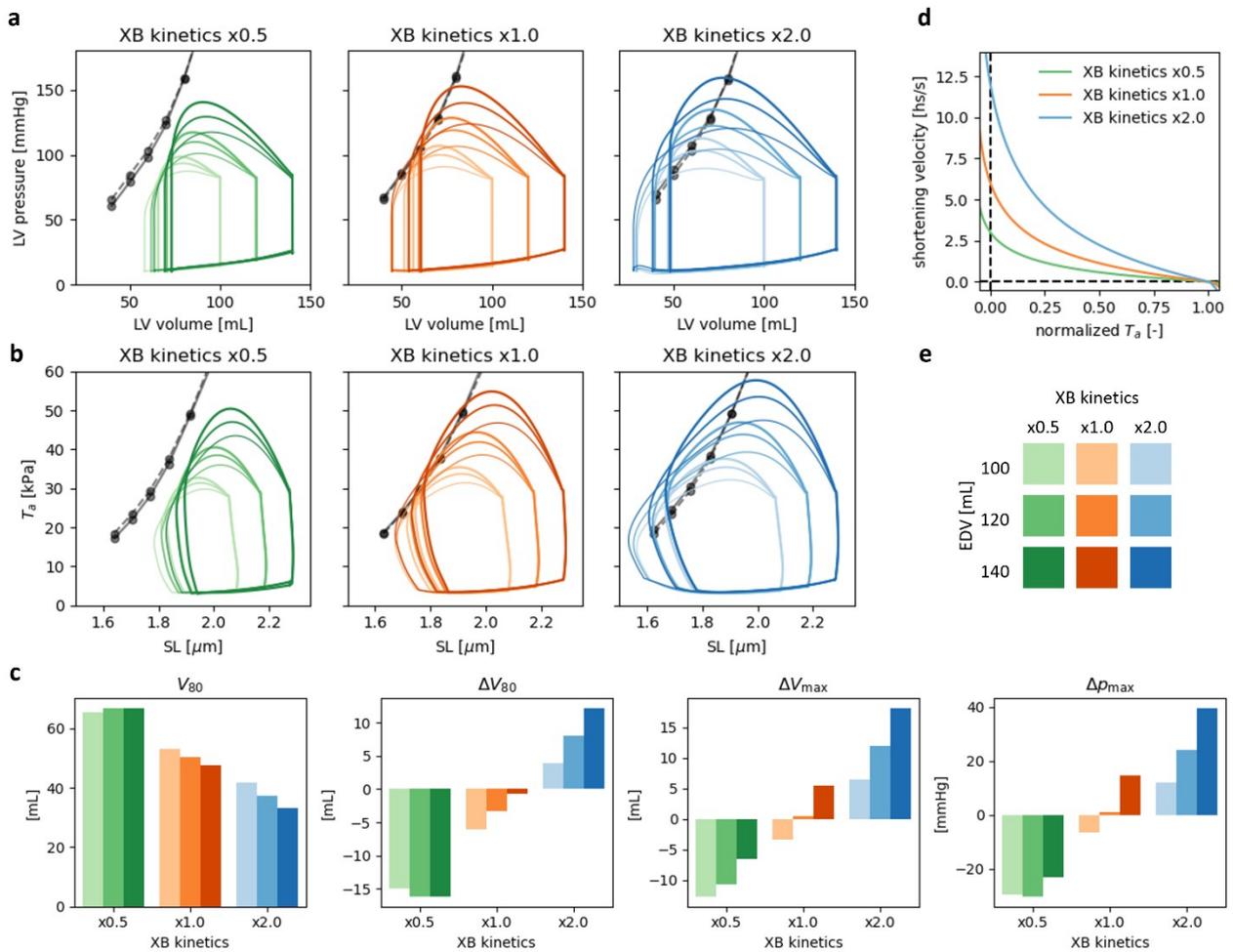

**Figure 3: Left ventricle model results for changes in XB kinetics.**

**a**: Pressure-volume loops obtained by varying XB kinetics (corresponding to line colour, see legend in e), EDV (corresponding to colour brightness, see legend in e), and arterial resistance (corresponding to line width, the thinner the line the lower the resistance – see text for more details). Black solid lines connecting the dots represent the isovolumetric pressure-volume curve associated with the XB kinetics considered. Black dashed lines instead represent the average pressure-volume curve among the three XB kinetics levels considered. **b**: Equivalent representation of a, showing space-averaged active tension versus sarcomere length instead of pressure and volume. **c**: dependence of $V_{80}$, $\Delta V_{80}$, $\Delta V_{\max}$ and $\Delta p_{\max}$ on XB kinetics and EDV (refer to colour legend in e). **d**: force-velocity relationship obtained from single-cell simulations for the three levels of XB kinetics considered here (velocity is measured as half sarcomere per second, hs/s). **e**: Colour legend.

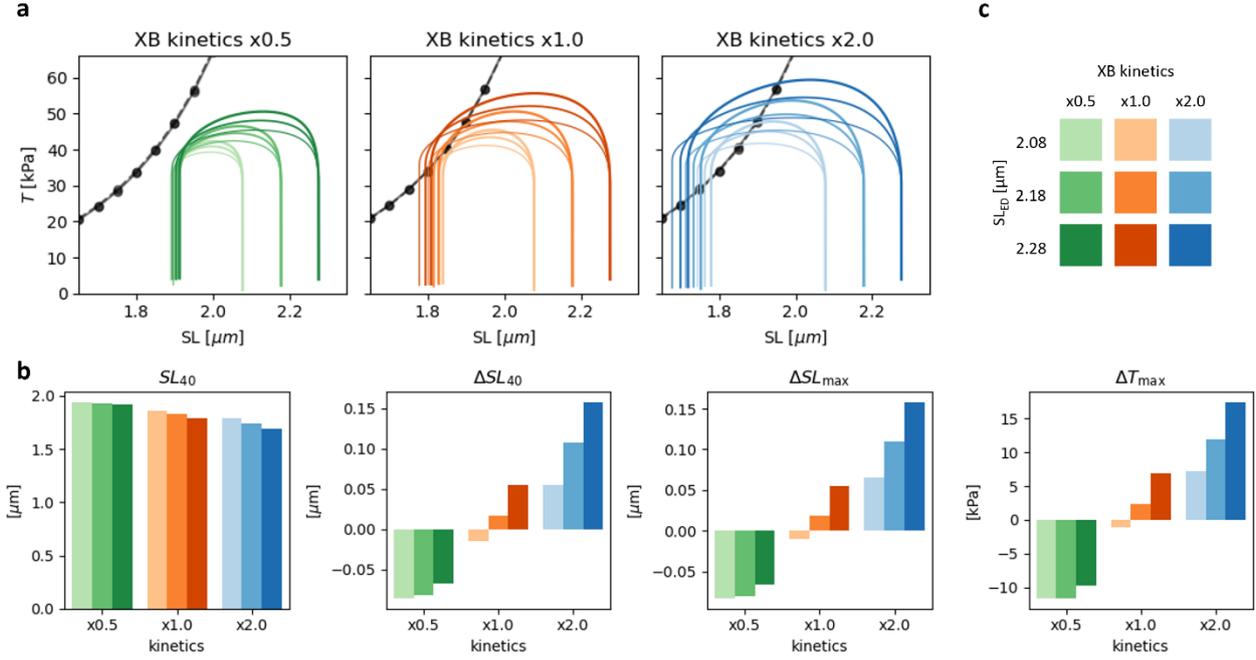

**Figure 4: isolated cardiac tissue model results for changes in XB kinetics.**

**a**: Tension-length curves obtained by varying XB kinetics (corresponding to line colour, see legend in e), $SL_{ED}$ (corresponding to colour brightness, see legend in e), and viscous coefficients (corresponding to line width, the thinner the line the lower the viscous damping – see text for more details). Black solid lines connecting the dots represent the isometric tension-length curve associated with the considered XB kinetics. Black dashed lines instead represent the average tension-length curve among the three XB kinetics levels considered. **b**: dependence of $SL_{40}$, $\Delta SL_{40}$, $\Delta SL_{max}$ and $\Delta T_{max}$ on XB kinetics and $SL_{ED}$ (see e for colour legend). **c**: colour legend.

## The impact of CaT

To investigate the role of calcium kinetics, we accelerated and decelerated the CaT used in baseline conditions by a factor of 20%. Note that the CaTs were adjusted uniformly, so that both the time-to-peak and the relaxation time were affected by the same factor. The resulting CaTs are shown in Figure 5d. We repeated the LV simulations, comparing the isovolumetric beats and the ejecting beats obtained with the different CaTs considered (see Figure 5). Firstly, we note that, not surprisingly, the longer the CaT, the higher the LV contractility, as evidenced by the pressure-volume loops and by the trend of both $V_{80}^{iso}$ and $V_{80}^{ejec}$. To quantify the presence of shortening activation/deactivation, we must compare the curves obtained with isovolumetric and ejecting beats, for a given CaT. This is done using the indices $\Delta V_{80}$, $\Delta V_{max}$ and $\Delta p_{max}$, that consistently indicate that a longer (or shorter) CaT leads to a shortening activation (or deactivation) effect. The same conclusions are drawn from the isolated tissue simulations (shown in Figure 6), by looking at $\Delta SL_{40}$, $\Delta SL_{max}$ and $\Delta T_{max}$. The impact of CaT duration was smaller when compared with that of XB kinetics reported above; however, it should be noted that in our investigation we considered milder variations in CaT duration than in XB kinetics, because too large an increase in CaT duration would prevent the LV from fully relaxing, thus rendering the overall simulations uninformative. A more direct comparison of the effects of XB kinetics and CaT duration can be found in the final section of the Results.

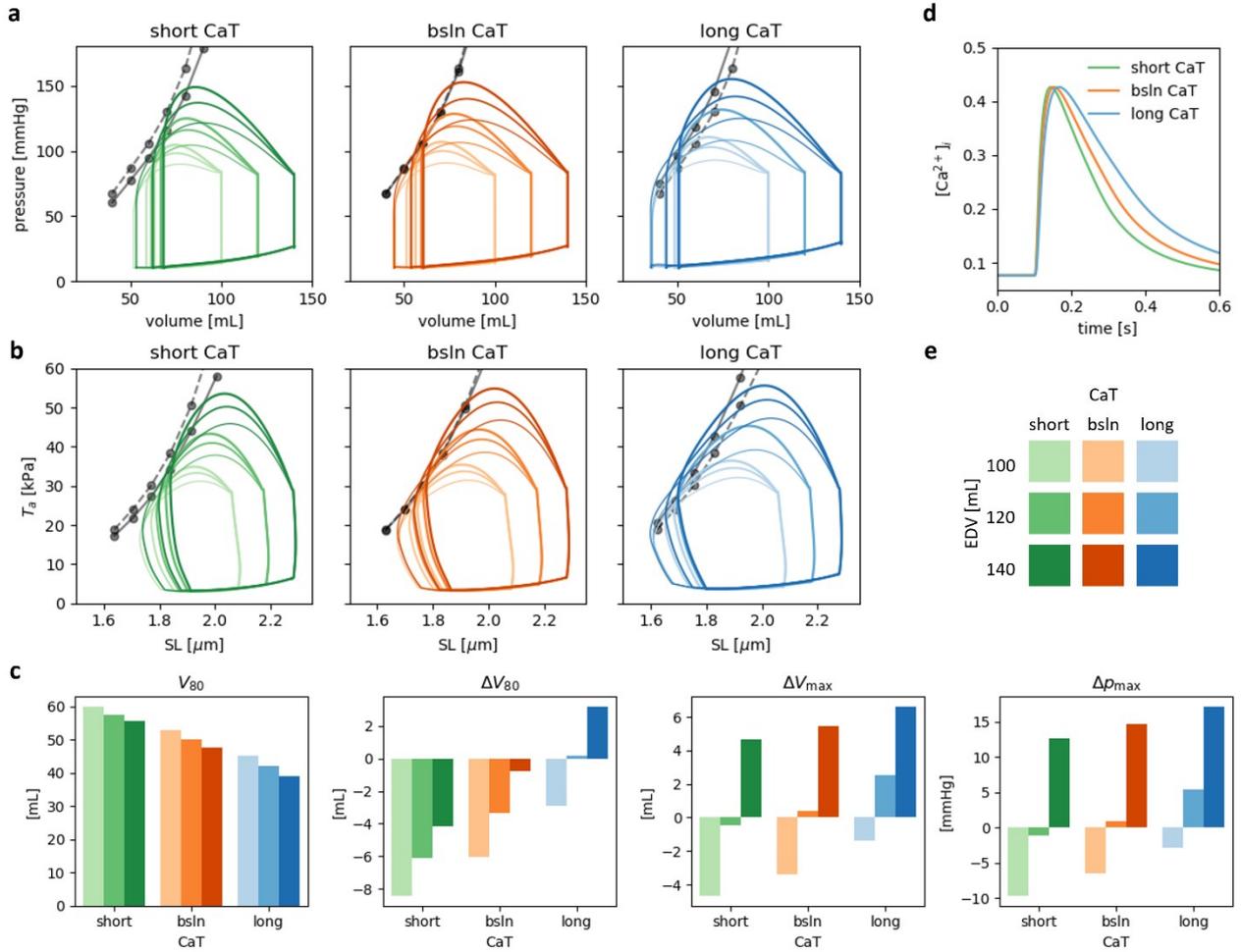

**Figure 5: Left ventricle model results for changes in CaT duration (at constant CaT amplitude).**

**a**: Pressure-volume loops obtained upon variations in CaT duration with constant CaT amplitude (corresponding to line colour, see legend in e), EDV (corresponding to colour brightness, see legend in e), and arterial resistance (corresponding to line width, the thinner the line the lower the resistance – see text for more details). Black solid lines connecting the dots represent the isovolumetric pressure-volume curve associated with the XB kinetics considered. Black dashed lines instead represent the average pressure-volume curve among the three CaTs considered. **b**: Equivalent representation of a, showing space-average active tension versus sarcomere length instead of pressure and volume. **c**: dependence of $V_{80}$, $\Delta V_{80}$, $\Delta V_{max}$ and $\Delta p_{max}$ on CaT duration and EDV (refer to colour legend in e). **d**: CaTs considered in the simulations reported in this figure. **e**: Colour legend.

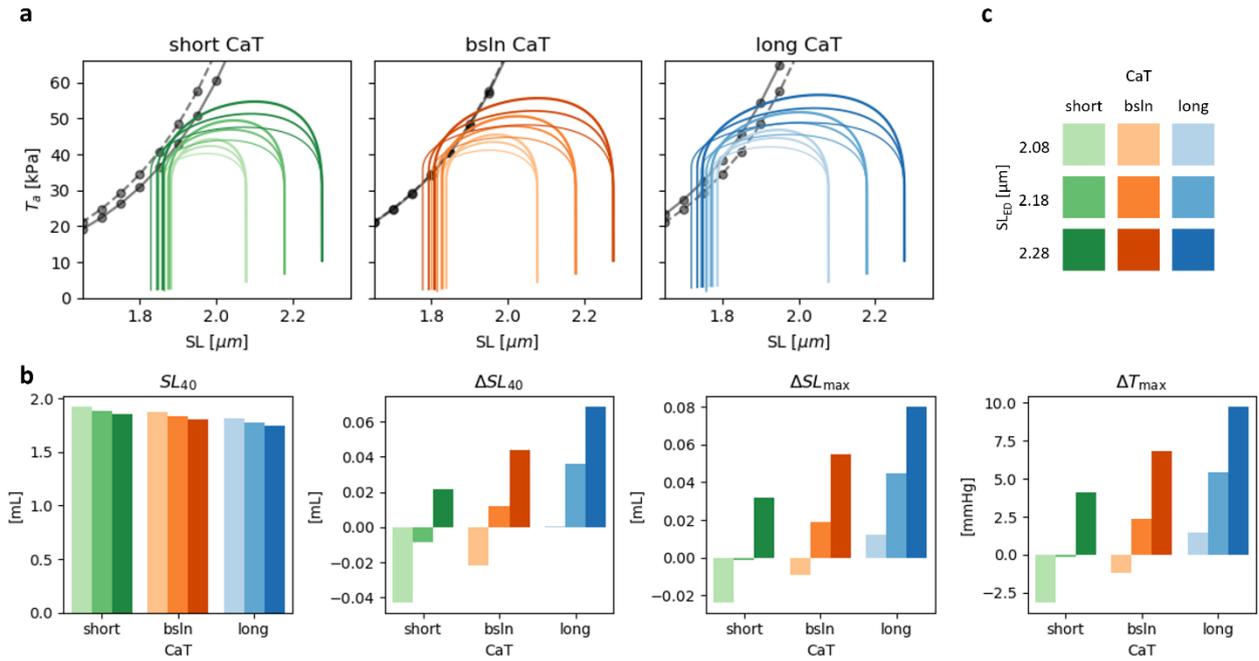

**Figure 6: isolated cardiac tissue model results for changes in CaT duration (at constant CaT amplitude).**

**a**: Tension-length curves obtained upon variations in CaT duration with constant CaT amplitude (corresponding to line colour, see legend in e), $SL_{ED}$ (corresponding to colour brightness, see legend in e), and viscous coefficients (corresponding to line width, the thinner the line the lower the viscous damping – see text for more details). Black solid lines connecting the dots represent the isometric tension-length curve associated with the considered XB kinetics. Black dashed lines instead represent the average tension-length curve among the three CaTs considered. **b**: dependence of $SL_{40}$, $\Delta SL_{40}$, $\Delta SL_{max}$ and $\Delta T_{max}$ on CaT duration and $SL_{ED}$ (see e for colour legend). **c**: colour legend.

However, as noted above, changing the CaT duration not only affects the pressure-volume loops of ejecting beats, but also shifts the isovolumetric ESPV curve, making it nontrivial to assess the role of CaT duration in determining the shortening effects on contractility. Therefore, in order to decouple the effect of CaT on the isovolumetric contractility from the effect of CaT on shortening activation/deactivation, we performed a second set of numerical tests, in which we corrected the CaT amplitude to compensate for the effects of CaT duration on isovolumetric contractility. Specifically, the amplitude of the longer CaT was reduced by 7% and the amplitude of the shorter CaT was increased by 7% (see Figure 7d). This correction was set to make the isovolumetric ESPV curves almost independent of the CaT considered, allowing a more direct comparison of the three cases in terms of the effect of shortening on contractility. It is noteworthy that the conclusions drawn above still hold even after the amplitude correction: the duration of CaT has a positive impact on contractility during shortening (see Figure 7).

In Figure 6 and Figure 8, we report the results obtained from isolated cardiac tissue simulations investigating the effects of CaT duration. We considered both the case of CaT duration modulation with constant amplitude (Figure 6), and CaT duration modulation with amplitude correction (Figure 8), by considering the same CaTs used for LV simulations. The results obtained were fully consistent with those obtained with LV simulations: the role played by CaT duration on contractility under isometric conditions and on the shortening effect on contractility was the same both at the tissue level and at the chamber level.

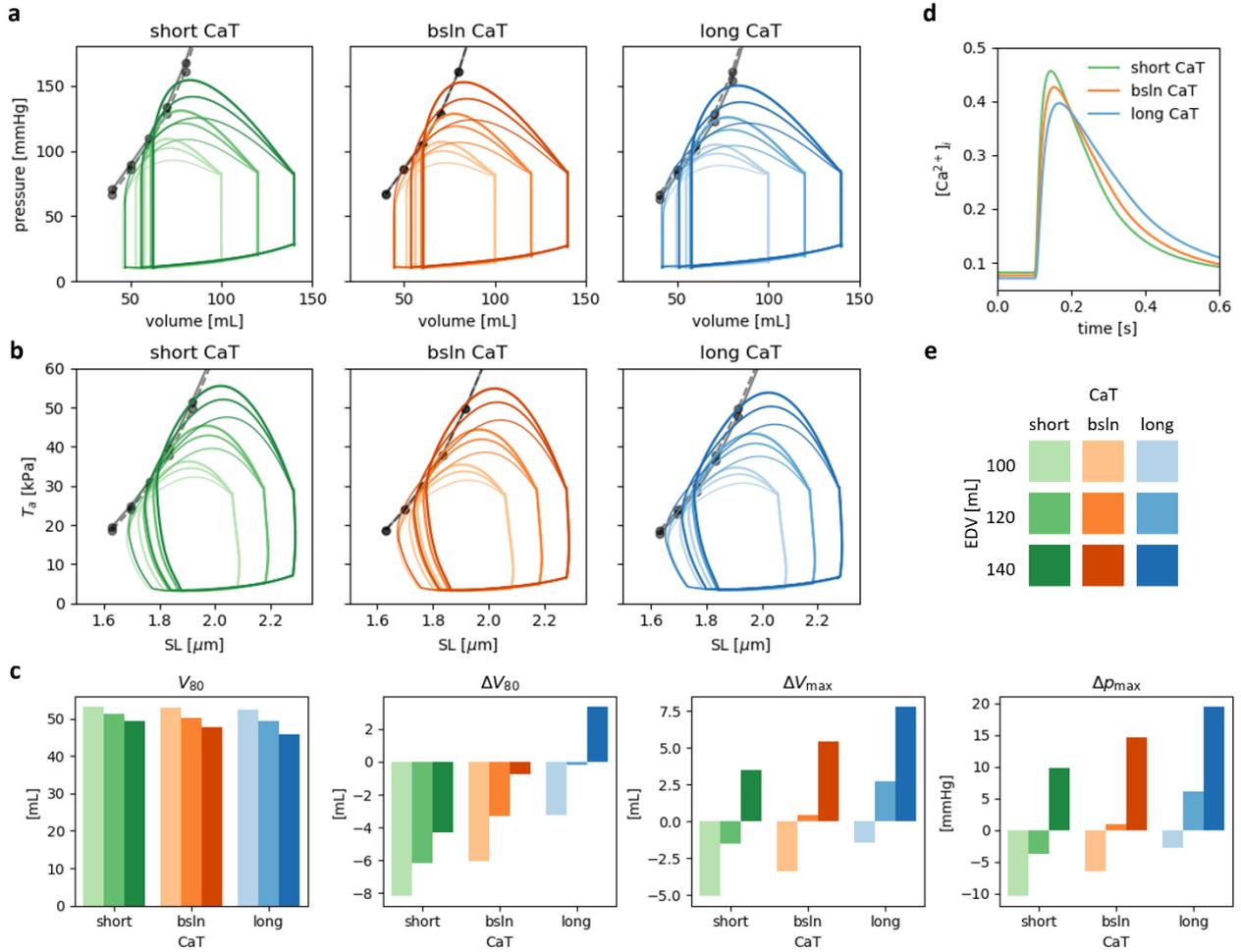

**Figure 7: Left ventricle model results for changes in CaT duration (with CaT amplitude correction).**

**a**: Pressure-volume loops obtained by varying the CaT duration with amplitude correction as described in the text (corresponding to line colour, see legend in e), EDV (corresponding to colour brightness, see legend in e), and arterial resistance (corresponding to line width, the thinner the line the lower the resistance – see text for more details). Black solid lines connecting the dots represent the isovolumetric pressure-volume curve associated with the XB kinetics considered. Black dashed lines instead represent the average pressure-volume curve among the three CaTs considered. **b**: Equivalent representation of a, showing space-average active tension versus sarcomere length instead of pressure and volume. **c**: dependence of $V_{80}$, $\Delta V_{80}$, $\Delta V_{\max}$ and $\Delta p_{\max}$ on CaT duration and EDV (refer to colour legend in e). **d**: CaTs considered in the simulations reported in this figure. **e**: Colour legend.

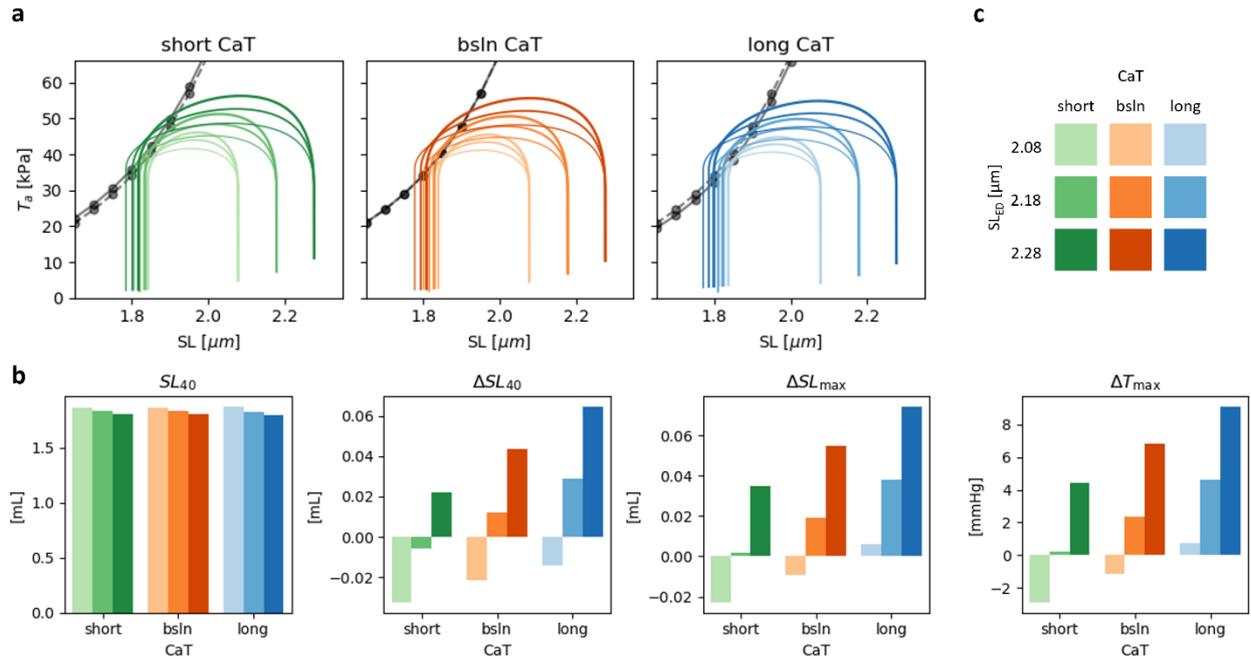

**Figure 8: Isolated cardiac tissue model results for changes in CaT duration (with CaT amplitude correction).**

**a**: Tension-length curves obtained by varying the CaT duration with amplitude correction as described in the text (corresponding to line colour, see legend in e), $SL_{ED}$ (corresponding to colour brightness, see legend in e), and viscous coefficients (corresponding to line width, the thinner the line the lower the viscous damping – see text for more details). Black solid lines connecting the dots represent the isometric tension-length curve associated with the considered XB kinetics. Black dashed lines instead represent the average tension-length curve among the three CaTs considered. **b**: dependence of $SL_{40}$, $\Delta SL_{40}$, $\Delta SL_{\max}$ and $\Delta T_{\max}$ on CaT duration and $SL_{ED}$ (see e for colour legend). **c**: colour legend.

## Deeper investigation of microscale variables

Building on the above results suggesting that the shortening effects on contractility are intrinsic to the cardiac tissue, we used the computational model of active force generation to gain mechanistic insight into the underlying cellular basis.

First, we considered the isolated tissue model, and we performed shortening twitch simulations by varying the XB kinetics. To allow for a finer investigation than that carried out in Figures 3 and 4, we considered milder variations (namely ±20%). We then performed an isometric simulation by choosing an SL value such that the isometric and shortening tension-length curves touched at the isometric peak. We then compared the transient of SL, of $T_a$, and of the regulatory units (RU) activation, i.e. the fraction of regulatory units that are in the unblocked state (i.e. they allow XB attachment to the associated actin units). The comparison is shown in Figure 9a. It is noteworthy that RU activation was significantly greater for shortening twitches (with a peak twice as large on average) than for isometric twitches. Increased XB kinetics had a (small) negative effect on the RU activation peak, and – in agreement with the above results – a more pronounced positive impact on the tension peak and on maximum shortening. In addition, increasing the XB kinetics slightly decreased the duration of both the RU activation and $T_a$ transients.

We performed a similar analysis to gain mechanistic insight into the impact of CaT duration. We performed three shortening twitch simulations, with baseline CaT and with 20% accelerated/decelerated CaTs, where we corrected the amplitude to achieve a constant tension peak. We compared the results with an isometric twitch, where SL was chosen so that the tension peak intersected the baseline tension-length curve (see Figure 9b). Again, the shortening twitches achieved a significantly larger RU activation peak than the isometric twitch. Increasing the CaT duration resulted in increased shortening, increased duration of both RU activation and $T_a$ transients, and a slight reduction in the peak of RU activation.

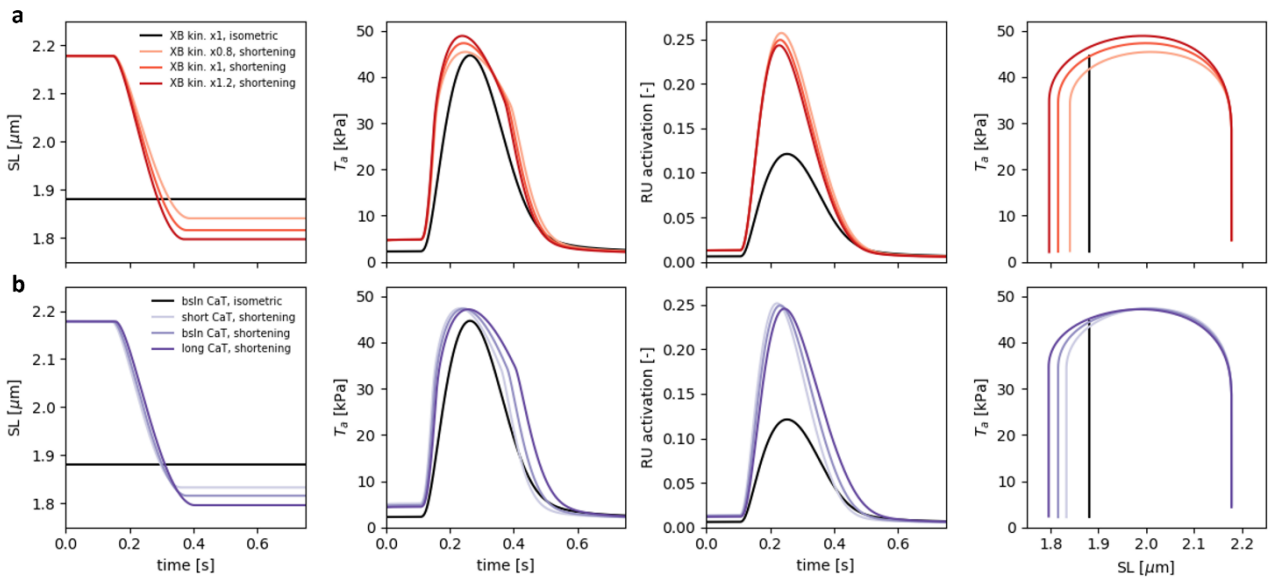

**Figure 9: Deeper investigation of microscale variables**

**a**: Traces associated with an isometric twitch and three shortening twitches with ±20% variation in XB kinetics from baseline. From left to right: sarcomere length; active tension; regulatory unit (RU) activation (i.e., the fraction of regulatory units that are in the unblocked state); tension-length curves. **b**: Traces associated with one isometric twitch and three shortening twitches with 20% accelerated/decelerated CaTs with amplitude correction, as described in the text.

## DISCUSSION

The computational results suggest that treating the ESPV relationship as a fixed curve is an oversimplification. Instead, this study suggests that, by connecting the end-systolic points obtained by varying the preload and afterload resistances, we do not obtain a single curve, but rather, in partial agreement with [14], a region within the pressure-volume plane. Furthermore, the amplitude of this region may depend on factors such as XB kinetics and CaT. Under baseline conditions, the spread of the curves is limited, so that the ESPV region effectively resembles a single curve if the resolution is insufficient. This may explain the traditional view of the ESPV relationship as a fixed curve. However, when XB kinetics are accelerated, the model predicts a significantly wider ESPV region, where a single curve approximation would indeed be an oversimplification.

Furthermore, this computational study shows that the mechanical mode of contraction plays a role in shaping the ESPV relationship. Under baseline conditions, the impact was negligible and dependent on the preload, with a small positive effect of ejection for large EDV and a small negative effect for small EDV. However, when variations from the baseline conditions were introduced, significant variations in the ESPV relationship of the ejecting beats were observed compared to the isovolumetric curve. This explains the observed variability of the pressure-volume loops, which can either exceed or remain below the isovolumetric curve, depending on the specific mechanical conditions and contraction modes, thus reconciling the apparently contradictory findings in the literature. Specifically, either an acceleration of the XB kinetics or a deceleration of the CaT leads to an enhancement of generated force during shortening, causing the ejecting ESPV curves to exceed the isovolumetric curve. Conversely, either a deceleration of XB kinetics or an acceleration of CaT resulted in ESPV curves below the isovolumetric curve. These findings highlight the complexity of cardiac muscle behaviour and show that contractility is not only defined by intrinsic properties but also by the temporal sequence of activation.

Remarkably, despite their simplicity, the isolated tissue simulations showed the same trends as the LV simulations, supporting the hypothesis that the impact of shortening on activation is due to mechanisms that are intrinsic to the tissue, rather than a property of the ventricular chamber. Furthermore, both shortening activation and deactivation were obtained as emergent phenomena of the model, as we did not introduce any artificial mechanism to modulate inotropy following the contraction mode. As discussed below, the two ingredients that were crucial to reproduce both phenomena were just the length-dependent apparent calcium sensitivity of troponin and the force-velocity relationship.

Motivated by this view, we used isolated tissue simulations to elucidate the mechanisms governing the effect of shortening on contractility. In fact, by simplifying the setup to its essentials, the comparisons shown in Figure 9 allows to gain insight into the origin of the phenomena observed above.

Let us first compare isometric and shortening twitches under baseline conditions (namely the black and coloured lines in Figure 9a). In the shortening case, activation starts at a higher SL than in the isometric case, and, due to the higher calcium-sensitivity associated with higher SL, the RU activation increases more steeply. The sarcomere then shortens to, and even falls below, the length assumed in the isometric case. This results into a lower calcium-sensitivity than in the isometric case, but the peak RU activation is still higher, since most of the thin filament activation process occurred at higher SLs, and thus with a troponin that was more able to bind calcium. This explains the higher RU activation obtained in the shortening case (coloured lines). However, looking at the plot of $T_a$, we can see that doubling the RU activation does not result in a similar increase in the force generated. The reason for this is that, due to the force-velocity relationship, the force is greatly reduced by shortening. This partially compensates for the increased activation of the thin filament. The conclusion is that the peak force obtained during shortening twitches is the net result of two opposing mechanisms that balance each other: on the one hand, an increased activation of the thin filament, which maintains the "memory" of the higher SL values present during the activation phase; on the other hand, the depressing effect on force due to the speed at which the sarcomere shortens. When the scales of this balance tip to one side or the other, we observe the two phenomena of shortening activation and deactivation described in the literature and also observed in the simulations reported in this paper.

With this in mind, let us now consider the effect of changes in the kinetics of XBs. It is well known that the higher the kinetic rates of XBs, the greater their ability to cycle and sustain high shortening velocities without excessive force reduction. Therefore, an increase (or decrease) in XB kinetics reduces (or accentuates) the effect of shortening velocity on the force generated (as also shown in Figure 3d). As a result, the second of the two compensatory mechanisms is weakened (or enhanced), leading to an increase (or decrease) in contractility. This is confirmed by the plots of $T_a$ in Figure 9a. Consequently, increased force generation leads to increased shortening, which in turn – due to the impact of SL on calcium-sensitivity – leads to an, albeit slightly, reduced RU activation, as shown in Figure 9a. However, this negative feedback loop does not cancel out the positive effect on $T_a$, as it is a secondary phenomenon.

Finally, in Figure 9b we examined the effect of changes in CaT duration. In this case, as shown in the figure, the two compensatory mechanisms are not significantly affected. However, because CaT is longer, for the same peak force (due to CaT amplitude modulation), a longer $T_a$ duration results in more tissue shortening. From another perspective, slowing down the kinetics of CaT has the effect of increasing the relative kinetics of XBs compared to CaT, leading to a shortening activation effect due to the mechanism described above.

We now address the potential translational significance of this work. It is noteworthy that both an acceleration of XB kinetics and a prolongation of CaT occur in hypertrophic cardiomyopathy (HCM), especially if associated with mutations in myosin binding protein C [49] or beta myosin [50], [51]. The model simulations presented in this paper indicate that this would result in an enhancement of the mechanical performance of ejecting beats, leading to a situation where the ESPV relationship of ejecting beats exceeds the isovolumetric curve. This may contribute to justify the description of HCM as a "hypercontractile" condition. The mutation-driven increase in XB kinetics is a hallmark of HCM and has never been reported in secondary hypertrophy or in heart failure (HF). Conversely, a slowing of CaT decay and, more generally CaT prolongation is a very common feature in HF of various aetiologies (e.g. ischaemic, hypertensive, and valvular defects). Based on our simulations, we can speculate that in HF a prolonged CaT may be an adaptive mechanism to maintain the stroke volume, at any given afterload level, despite the reduced force generating capacity of the myocardium (e.g. reduced contractility, reduced myofilament density, myocardial disarray).

Another point to consider is the relative expression of alpha and beta myosin isoforms in human myocardium. The rate of XB cycling is 2.5-3 times higher with alpha myosin than with beta myosin (for review see [52]). The relative expression of alpha and beta myosin heavy chains (encoded by the MYH6 and MYH7 genes, respectively) has been shown to be controlled by thyroid hormones (specifically 3,5,3'-triiodo-L-thyronine, T3) in rodents [53]. Although little is known about the regulation of the expression of the alpha-MHC gene in the human ventricles, the similarities observed in the 5' flanking region between the rat and human alpha-MHC genes suggest that the human gene may be regulated in a similar way [54]. The acceleration of XB kinetics resulting from a switch from beta to alpha myosin isoforms may enhance force generation during shortening, thereby contributing to the hyperdynamic conditions associated with hyperthyroidism.

A limitation of this study is that we have neglected the mechano-calcium feedback [55], [56], by which the free cytosolic calcium concentration is affected by the calcium-troponin binding process, potentially leading to differences in the CaT of isometric vs shortening twitches. Although this may alter the results on a quantitative level, we do not expect any significant changes on a qualitative level, and the mechanisms elucidated in this study would remain essentially unchanged. In addition, both isovolumetric and ejecting beats were simulated at steady state without considering cell mechanisms that may be involved in the slow force response to changes in initial cardiac muscle length [57], [58] or in the von Anrep effect [59], [60].

## REFERENCES

[1] H. Suga, 'Ventricular energetics', *Physiol Rev*, vol. 70, no. 2, pp. 247–277, Apr. 1990, doi: 10.1152/physrev.1990.70.2.247.
[2] O. Frank, 'Die grundform des arterielen pulses erste abhandlung: mathematische analyse', *Zeitschrift für Biologie*, vol. 37, pp. 483–526.
[3] K. Sagawa, R. K. Lie, and J. Schaefer, 'Translation of Otto Frank's paper "Die Grundform des Arteriellen Pulses" Zeitschrift für Biologie 37: 483-526 (1899)', *J Mol Cell Cardiol*, vol. 22, no. 3, pp. 253–254, Mar. 1990, doi: 10.1016/0022-2828(90)91459-k.
[4] J.-C. Han, D. Loiselle, A. Taberner, and K. Tran, 'Re-visiting the Frank-Starling nexus', *Progress in Biophysics and Molecular Biology*, vol. 159, pp. 10–21, Jan. 2021, doi: 10.1016/j.pbiomolbio.2020.04.003.
[5] H. Suga, K. Sagawa, and A. A. Shoukas, 'Load independence of the instantaneous pressure-volume ratio of the canine left ventricle and effects of epinephrine and heart rate on the ratio', *Circ Res*, vol. 32, no. 3, pp. 314–322, Mar. 1973, doi: 10.1161/01.res.32.3.314.
[6] K. Sagawa, L. Maughan, H. Suga, and K. Sunagawa, *Cardiac Contraction and the Pressure Volume Relationship*. New York Oxford University Press, 1988.

[7] K. A. Edman, 'Mechanical deactivation induced by active shortening in isolated muscle fibres of the frog.', *The Journal of Physiology*, vol. 246, no. 1, pp. 255–275, 1975, doi: 10.1113/jphysiol.1975.sp010889.

[8] J. K. Leach, A. J. Brady, B. J. Skipper, and D. L. Millis, 'Effects of active shortening on tension development of rabbit papillary muscle', *American Journal of Physiology-Heart and Circulatory Physiology*, vol. 238, no. 1, pp. H8–H13, Jan. 1980, doi: 10.1152/ajpheart.1980.238.1.H8.

[9] N. Westerhof and G. Elzinga, 'The apparent source resistance of heart and muscle', *Ann Biomed Eng*, vol. 6, no. 1, pp. 16–32, Mar. 1978, doi: 10.1007/BF02584529.

[10] S. G. Shroff, J. S. Janicki, and K. T. Weber, 'Evidence and quantitation of left ventricular systolic resistance', *Am J Physiol*, vol. 249, no. 2 Pt 2, pp. H358-370, Aug. 1985, doi: 10.1152/ajpheart.1985.249.2.H358.

[11] J.-C. Han, T. Pham, A. J. Taberner, D. S. Loiselle, and K. Tran, 'Solving a century-old conundrum underlying cardiac force-length relations', *American Journal of Physiology-Heart and Circulatory Physiology*, vol. 316, no. 4, pp. H781–H793, Apr. 2019, doi: 10.1152/ajpheart.00763.2018.

[12] D. Burkhoff, P. P. De Tombe, and W. C. Hunter, 'Impact of ejection on magnitude and time course of ventricular pressure-generating capacity', *American Journal of Physiology-Heart and Circulatory Physiology*, vol. 265, no. 3, pp. H899–H909, Sep. 1993, doi: 10.1152/ajpheart.1993.265.3.H899.

[13] Y. Igarashi, C. P. Cheng, and W. C. Little, 'Left ventricular ejection activation in the in situ heart', *American Journal of Physiology-Heart and Circulatory Physiology*, vol. 260, no. 5, pp. H1495–H1500, May 1991, doi: 10.1152/ajpheart.1991.260.5.H1495.

[14] P. P. De Tombe and W. C. Little, 'Inotropic effects of ejection are myocardial properties', *American Journal of Physiology-Heart and Circulatory Physiology*, vol. 266, no. 3, pp. H1202–H1213, Mar. 1994, doi: 10.1152/ajpheart.1994.266.3.H1202.

[15] A. Landesberg, 'End-systolic pressure-volume relationship and intracellular control of contraction', *American Journal of Physiology-Heart and Circulatory Physiology*, vol. 270, no. 1, pp. H338–H349, Jan. 1996, doi: 10.1152/ajpheart.1996.270.1.H338.

[16] J. Shimizu, K. Todaka, and D. Burkhoff, 'Load dependence of ventricular performance explained by model of calcium-myofilament interactions', *American Journal of Physiology-Heart and Circulatory Physiology*, vol. 282, no. 3, pp. H1081–H1091, Mar. 2002, doi: 10.1152/ajpheart.00498.2001.

[17] G. Iribe, T. Kaneko, Y. Yamaguchi, and K. Naruse, 'Load dependency in force–length relations in isolated single cardiomyocytes', *Progress in Biophysics and Molecular Biology*, vol. 115, no. 2, pp. 103–114, Aug. 2014, doi: 10.1016/j.pbiomolbio.2014.06.005.

[18] M. Kalda, P. Peterson, and M. Vendelin, 'Cross-Bridge Group Ensembles Describing Cooperativity in Thermodynamically Consistent Way', *PLOS ONE*, vol. 10, no. 9, p. e0137438, Sep. 2015, doi: 10.1371/journal.pone.0137438.

[19] A. Pironet *et al.*, 'A multi-scale cardiovascular system model can account for the load-dependence of the end-systolic pressure-volume relationship', *BioMedical Engineering OnLine*, vol. 12, no. 1, p. 8, Jan. 2013, doi: 10.1186/1475-925X-12-8.

[20] E. J. Botcherby *et al.*, 'Fast Measurement of Sarcomere Length and Cell Orientation in Langendorff-Perfused Hearts Using Remote Focusing Microscopy', *Circulation Research*, vol. 113, no. 7, pp. 863–870, Sep. 2013, doi: 10.1161/CIRCRESAHA.113.301704.

[21] F. Kobirumaki-Shimozawa *et al.*, 'Nano-imaging of the beating mouse heart in vivo: Importance of sarcomere dynamics, as opposed to sarcomere length per se, in the regulation of cardiac function', *Journal of General Physiology*, vol. 147, no. 1, pp. 53–62, Jan. 2016, doi: 10.1085/jgp.201511484.

[22] A. Santiago *et al.*, 'Fully coupled fluid-electro-mechanical model of the human heart for supercomputers: f-e-m model of the heart for supercomputers', *Int J Numer Meth Biomed Engng*, vol. 34, no. 12, p. e3140, Dec. 2018, doi: 10.1002/cnm.3140.

[23] C. M. Augustin *et al.*, 'Anatomically accurate high resolution modeling of human whole heart electromechanics: A strongly scalable algebraic multigrid solver method for nonlinear deformation', *Journal of Computational Physics*, vol. 305, pp. 622–646, Jan. 2016, doi: 10.1016/j.jcp.2015.10.045.

[24] M. R. Pfaller *et al.*, 'The importance of the pericardium for cardiac biomechanics: from physiology to computational modeling', *Biomech Model Mechanobiol*, vol. 18, no. 2, pp. 503–529, Apr. 2019, doi: 10.1007/s10237-018-1098-4.

[25] M. Strocchi, 'Simulating ventricular systolic motion in a four-chamber heart model with spatially varying robin boundary conditions to model the effect of the pericardium', *Journal of Biomechanics*, p. 9, 2020.

[26] T. Fritz, C. Wieners, G. Seemann, H. Steen, and O. Dössel, 'Simulation of the contraction of the ventricles in a human heart model including atria and pericardium', *Biomech Model Mechanobiol*, vol. 13, no. 3, pp. 627–641, Jun. 2014, doi: 10.1007/s10237-013-0523-y.

[27] F. Regazzoni and A. Quarteroni, 'An oscillation-free fully staggered algorithm for velocity-dependent active models of cardiac mechanics', *Computer Methods in Applied Mechanics and Engineering*, vol. 373, p. 113506, Jan. 2021, doi: 10.1016/j.cma.2020.113506.

[28] F. Regazzoni, M. Salvador, P. C. Africa, M. Fedele, L. Dedè, and A. Quarteroni, 'A cardiac electromechanical model coupled with a lumped-parameter model for closed-loop blood circulation', *Journal of Computational Physics*, vol. 457, p. 111083, May 2022, doi: 10.1016/j.jcp.2022.111083.

[29] M. Fedele *et al.*, 'A comprehensive and biophysically detailed computational model of the whole human heart electromechanics', *Computer Methods in Applied Mechanics and Engineering*, vol. 410, p. 115983, May 2023, doi: 10.1016/j.cma.2023.115983.

[30] J. Tomek *et al.*, 'Development, calibration, and validation of a novel human ventricular myocyte model in health, disease, and drug block', *eLife*, vol. 8, p. e48890, Dec. 2019, doi: 10.7554/eLife.48890.

[31] F. Regazzoni, L. Dedè, and A. Quarteroni, 'Biophysically detailed mathematical models of multiscale cardiac active mechanics', *PLoS Comput Biol*, vol. 16, no. 10, p. e1008294, Oct. 2020, doi: 10.1371/journal.pcbi.1008294.

[32] J. J. Rice and P. P. de Tombe, 'Approaches to modeling crossbridges and calcium-dependent activation in cardiac muscle', *Progress in Biophysics and Molecular Biology*, vol. 85, no. 2–3, pp. 179–195, Jun. 2004, doi: 10.1016/j.pbiomolbio.2004.01.011.

[33] F. Regazzoni, L. Dedè, and A. Quarteroni, 'Active contraction of cardiac cells: a reduced model for sarcomere dynamics with cooperative interactions', *Biomech Model Mechanobiol*, vol. 17, no. 6, pp. 1663–1686, Dec. 2018, doi: 10.1007/s10237-018-1049-0.

[34] S. Stella, F. Regazzoni, C. Vergara, L. Dedé, and A. Quarteroni, 'A fast cardiac electromechanics model coupling the Eikonal and the nonlinear mechanics equations', *Math. Models Methods Appl. Sci.*, vol. 32, no. 08, pp. 1531–1556, Jul. 2022, doi: 10.1142/S021820252250035X.

[35] P. Colli Franzone, L. Guerri, and S. Rovida, 'Wavefront propagation in an activation model of the anisotropic cardiac tissue: asymptotic analysis and numerical simulations', *J. Math. Biol.*, vol. 28, no. 2, pp. 121–176, Feb. 1990, doi: 10.1007/BF00163143.

[36] P. Colli Franzone, L. F. Pavarino, and S. Scacchi, *Mathematical Cardiac Electrophysiology*, vol. 13. in MS&A, vol. 13. Cham: Springer International Publishing, 2014. doi: 10.1007/978-3-319-04801-7.

[37] M. E. Gurtin, E. Fried, and L. Anand, 'The Mechanics and Thermodynamics of Continua', p. 718.

[38] T. P. Usyk, I. J. LeGrice, and A. D. McCulloch, 'Computational model of three-dimensional cardiac electromechanics', *Comput Visual Sci*, vol. 4, no. 4, pp. 249–257, Jul. 2002, doi: 10.1007/s00791-002-0081-9.

[39] Zygote, 'Zygote solid 3D male anatomy collection generation II develompent report', 2014.

[40] N. A. Barnafi, F. Regazzoni, and D. Riccobelli, 'Reconstructing relaxed configurations in elastic bodies: Mathematical formulations and numerical methods for cardiac modeling', *Computer Methods in Applied Mechanics and Engineering*, vol. 423, p. 116845, Apr. 2024, doi: 10.1016/j.cma.2024.116845.

[41] J. D. Bayer, R. C. Blake, G. Plank, and N. A. Trayanova, 'A Novel Rule-Based Algorithm for Assigning Myocardial Fiber Orientation to Computational Heart Models', *Ann Biomed Eng*, vol. 40, no. 10, pp. 2243–2254, Oct. 2012, doi: 10.1007/s10439-012-0593-5.

[42] R. Piersanti *et al.*, 'Modeling cardiac muscle fibers in ventricular and atrial electrophysiology simulations', *Computer Methods in Applied Mechanics and Engineering*, vol. 373, p. 113468, Jan. 2021, doi: 10.1016/j.cma.2020.113468.

[43] F. Regazzoni, L. Dedè, and A. Quarteroni, 'Machine learning of multiscale active force generation models for the efficient simulation of cardiac electromechanics', *Computer Methods in Applied Mechanics and Engineering*, vol. 370, p. 113268, Oct. 2020, doi: 10.1016/j.cma.2020.113268.

[44] S. Land, S.-J. Park-Holohan, N. P. Smith, C. G. dos Remedios, J. C. Kentish, and S. A. Niederer, 'A model of cardiac contraction based on novel measurements of tension development in human cardiomyocytes', *Journal of Molecular and Cellular Cardiology*, vol. 106, pp. 68–83, May 2017, doi: 10.1016/j.yjmcc.2017.03.008.

[45] M. B. Bastos *et al.*, 'Invasive left ventricle pressure–volume analysis: overview and practical clinical implications', *European Heart Journal*, vol. 41, no. 12, pp. 1286–1297, Mar. 2020, doi: 10.1093/eurheartj/ehz552.

[46] F. Regazzoni, 'Mathematical Modeling and Machine Learning for the Numerical Simulation of Cardiac Electromechanics', Politecnico di Milano, 2020.

[47] *lifex*. [Online]. Available: https://lifex.gitlab.io/

[48] P. C. Africa, 'lifex: A flexible, high performance library for the numerical solution of complex finite element problems', *SoftwareX*, vol. 20, p. 101252, Dec. 2022, doi: 10.1016/j.softx.2022.101252.

[49] J. M. Pioner *et al.*, 'Slower Calcium Handling Balances Faster Cross-Bridge Cycling in Human MYBPC3 HCM', *Circulation Research*, vol. 132, no. 5, p. 628, Feb. 2023, doi: 10.1161/CIRCRESAHA.122.321956.

[50] A. Belus *et al.*, 'The familial hypertrophic cardiomyopathy-associated myosin mutation R403Q accelerates tension generation and relaxation of human cardiac myofibrils', *J Physiol*, vol. 586, no. 15, pp. 3639–3644, Aug. 2008, doi: 10.1113/jphysiol.2008.155952.

[51] C. Poggesi and C. Y. Ho, 'Muscle dysfunction in hypertrophic cardiomyopathy: What is needed to move to translation?', *Journal of muscle research and cell motility*, vol. 35, no. 1, p. 37, Feb. 2014, doi: 10.1007/s10974-014-9374-0.

[52] J. Walklate, C. Ferrantini, C. A. Johnson, C. Tesi, C. Poggesi, and M. A. Geeves, 'Alpha and beta myosin isoforms and human atrial and ventricular contraction', *Cell. Mol. Life Sci.*, vol. 78, no. 23, pp. 7309–7337, Dec. 2021, doi: 10.1007/s00018-021-03971-y.

[53] T. A. Gustafson, B. E. Markham, and E. Morkin, 'Effects of thyroid hormone on alpha-actin and myosin heavy chain gene expression in cardiac and skeletal muscles of the rat: measurement of mRNA content using synthetic oligonucleotide probes.', *Circulation Research*, vol. 59, no. 2, pp. 194–201, Aug. 1986, doi: 10.1161/01.RES.59.2.194.

[54] R. W. Tsika, J. J. Bahl, L. A. Leinwand, and E. Morkin, 'Thyroid hormone regulates expression of a transfected human alpha-myosin heavy-chain fusion gene in fetal rat heart cells.', *Proc. Natl. Acad. Sci. U.S.A.*, vol. 87, no. 1, pp. 379–383, Jan. 1990, doi: 10.1073/pnas.87.1.379.

[55] F. Levrero-Florencio *et al.*, 'Sensitivity analysis of a strongly-coupled human-based electromechanical cardiac model: Effect of mechanical parameters on physiologically relevant biomarkers', *Computer Methods in Applied Mechanics and Engineering*, vol. 361, p. 112762, Apr. 2020, doi: 10.1016/j.cma.2019.112762.

[56] F. Mazhar *et al.*, 'A detailed mathematical model of the human atrial cardiomyocyte: integration of electrophysiology and cardiomechanics', *The Journal of Physiology*, p. JP283974, Aug. 2023, doi: 10.1113/JP283974.

[57] W. W. Parmley and L. Chuck, 'Length-dependent changes in myocardial contractile state', *Am J Physiol*, vol. 224, no. 5, pp. 1195–1199, May 1973, doi: 10.1152/ajplegacy.1973.224.5.1195.

[58] J. Dowrick *et al.*, 'The Slow Force Response to Stretch: Controversy and Contradictions', *Acta Physiologica*, vol. 226, p. e13250, Jan. 2019, doi: 10.1111/apha.13250.

[59] G. von Anrep, 'On the part played by the suprarenals in the normal vascular reactions of the body', *The Journal of Physiology*, vol. 45, no. 5, p. 307, Dec. 1912, doi: 10.1113/jphysiol.1912.sp001553.

[60] V. Sequeira, C. Maack, G.-H. Reil, and J.-C. Reil, 'Exploring the Connection Between Relaxed Myosin States and the Anrep Effect', *Circulation Research*, vol. 134, no. 1, pp. 117–134, Jan. 2024, doi: 10.1161/CIRCRESAHA.123.323173.

## ADDITIONAL INFORMATION

### Data Availability Statement

This study does not involve any experimental data, as it is solely based on numerical simulations.

### Competing Interests

None of the authors has any conflicts of interests.

### Author contributions

F.R.: Conceptualization, Formal analysis, Funding acquisition, Investigation, Methodology, Software, Visualization, Writing – original draft, Writing – review & editing.

C.P.: Conceptualization, Funding acquisition, Methodology, Writing – original draft, Writing – review & editing.

C.F.: Conceptualization, Funding acquisition, Methodology, Writing – original draft, Writing – review & editing.

### Funding

F.R. has received funding from the project PRIN2022, MUR, Italy, 2023-2025, P2022N5ZNP "SIDDMs: shape-informed data-driven models for parametrized PDEs, with application to computational cardiology", founded by the European Union (Next Generation EU, Mission 4 Component 2). The present research is part of the activities of "Dipartimento di Eccellenza 2023–2027", MUR, Italy, Dipartimento di Matematica, Politecnico di Milano. C.F. and C.P. have received funding from the European Union's Horizon 2020 under grant agreement No 952166 REPAIR.

### Acknowledgements

F.R. acknowledges his membership to INdAM GNCS - Gruppo Nazionale per il Calcolo Scientifico (National Group for Scientific Computing, Italy).